\documentclass[12pt]{article}

\usepackage{graphicx}

\newcommand{\zpr}{\mbox{$Z^{\prime}$}}
\newcommand{\upr}{\mbox{$U(1)^{\prime}$}}

\newcommand{\x}{\mbox{$\times$}}

\begin{document}
\DeclareGraphicsExtensions{.eps,.jpg,.pdf,.mps,.png}
\begin{flushright}
\baselineskip=12pt
UPR-1076-T \\
\end{flushright}

\begin{center}
\vglue 1.5cm

{\Large\bf $Z'$ Discovery Limits For Supersymmetric $E_6$ Models}
\vglue 2.0cm {\Large Junhai Kang$^a$, Paul Langacker$^a$ } \vglue
1cm { $^a$ Department of Physics and Astronomy,
University of Pennsylvania, \\Philadelphia, PA 19104-6396, USA \\
}
\end{center}

\vglue 1.0cm
\begin{abstract}
We study how the exotic particles and supersymmetric partners
would affect the discovery limit at the Tevatron and LHC for
neutral gauge bosons in generic $E_6$ models. We examine the $Z'$
decay in the extreme case that all of the particles are massless,
then consider how the masses of non-standard model particles will
affect the discovery limit. We also calculate the discovery limit
for a supersymmetric $E_6$ model with a secluded sector as a
concrete example of a model with a definite set of exotic
particles. Its discovery limit is small compared with other $E_6$
models due to the $U'(1)$ charge assignment.

\end{abstract}

\vspace{0.5cm}
\begin{flushleft}
\baselineskip=12pt
\today\\
\end{flushleft}
\newpage
\baselineskip=14pt

\section{Introduction}

While the standard model (SM) has been precisely tested by experiments,
it is believed to be only a low energy effective description of nature.
Extended gauge symmetries and/or extra gauge bosons appear
in many extensions of the standard model, such as left-right
symmetric models~\cite{lrm}, superstring motivated models~\cite{string}, GUT~(grand
unification theory)~\cite{review}, little Higgs models~\cite{lHiggs}, large extra
dimensions~\cite{led},
and dynamical symmetry breaking~\cite{DSB}.
They
are good candidates for new physics at future colliders. To date,
extra gauge bosons have not been observed, setting constraints on
various models.

In this paper, we focus on extra neutral gauge bosons~\cite{reviews}. The direct
search for $Z'$ requires the collider energy to be high enough to
produce the $Z'$ and the signal to be distinguished from the
standard model background. The discovery limit for the $Z'$ mass
is model dependent, but it is convenient to give it
for some representative models, such as the SSM~(sequential
standard model), LRM~(left-right symmetric model), and $E_6$ models.
The CDF and DO
collaborations~\cite{explim} give direct $Z'$ search limits ranging from 560
GeV to 690 GeV for different specific models from their Run I
data from the non-observation of $\bar{p}p \rightarrow e^+e^-,~\mu^+\mu^-$.
These should be improved to $O$(800 GeV) from these and other decay modes
in Run II, 
and to several TeV at the LHC.
The discovery reach for various models at future hadron and
lepton colliders was studied in~\cite{capstick}-\cite{carena}.
There are also stringent indirect limits,
especially on $Z-Z'$ mixing,
from low energy, $Z$-pole, and LEP2 experiments~\cite{indirect1}-\cite{cheung}.

The model dependence arises in part because of the $U(1)'$ gauge
couplings and quark and lepton charges. However, $U(1)'$ models
necessarily imply new exotic fermions needed for anomaly
cancellations~\cite{erler}.
If some of these and/or some of the superpartners in
supersymmetric versions are light enough to be produced in $Z'$
decays, the leptonic branching ratios and discovery limits will be
affected. Some earlier discussions of the implications of exotics and superpartners on
the discovery limits are found in~\cite{durkin}-\cite{gherghetta}. In this paper, we study
these effects in detail for the example of generic $E_6$-motivated
\zpr\ models. The $E_6$ gauge group can be broken as
\begin{eqnarray}
E_6 \to\ SO(10) \x\ U(1)_{\psi} \to\ SU(5) \x\ U(1)_{\chi} \x\
U(1)_{\psi}~.~\,
\end{eqnarray}
The $U(1)_{\psi}$ and $U(1)_{\chi}$ charges for the $E_6$
fundamental representation ${\bf 27}$ are given in Table
\ref{E6charge}.
Each {\bf 27} includes one SM family, an exotic charge $-1/3$ $D$ quark and
its conjugate; two SM singlets $\overline{N}$ and $S_L$; and one pair of Higgs-like
doublets $H^{\prime}_{u,d}$. One pair can be associated with the MSSM
Higgs doublets, and the other two as exotics\footnote{It is sometimes
convenient, depending on the symmetries of the superpotential, to interpret the
exotic Higgs fields as exotic lepton doublets, which have the same gauge quantum
numbers.}.
As described below, we will also add some additional doublets and singlets from
${\bf 27}+\overline{\bf 27}$ pairs.

The \upr\ in consideration is one linear
combination of the $U(1)_{\chi}$ and $U(1)_{\psi}$
\begin{eqnarray}
Q^{\prime} &=& \cos\theta \ Q_{\chi} + \sin\theta \ Q_{\psi}~,~\,
\label{E6MIX}
\end{eqnarray}
parameterized by angle $\theta$. For simplicity, we assume that
the other $U(1)$ gauge symmetry from the orthogonal linear
combination of the $U(1)_{\chi}$ and $U(1)_{\psi}$ is absent or
broken at a high scale\footnote{We are not considering a full
$E_6$ grand unified theory, because a light \zpr\ would prevent a
large doublet-triplet splitting and lead to rapid proton decay.
Rather, it is convenient and conventional to simply consider the
\upr\ charge assignment and exotic particle content of $E_6$ as an
example of an anomaly free construction. Other examples, such as
superstring motivated models, typically have more complicated
exotic structures~\cite{string2}.}. $U(1)_{\eta}$ is a particular combination of
$U(1)_{\chi}$ and $U(1)_{\psi}$, i.e.,
 $\theta = 2 \pi - \tan^{-1} \sqrt{\frac{5}{3}}=1.71 \pi$.
It occurs in Calabi-Yau compactifications of the heterotic string
model if $E_6$ leads directly to a rank 5 group via the Wilson
line (Hosotani) mechanism. $U(1)_N$ is a special model used in
~\cite{Hambye:2000bn,neutralN}, in which the right-handed neutrino
decouples from the low energy $U(1)$, avoiding big bang nucleosynthesis constraints.
It corresponds to $\theta = \tan^{-1} \sqrt{15}\sim 0.42 \pi$.

\begin{table}[t]
\caption{Decomposition of the $E_6$ fundamental representation
${\bf 27}$ under $SO(10)$ and $SU(5)$, and their $U(1)_{\chi}$,
$U(1)_{\psi}$, $U(1)_{\eta}$ , secluded sector $U(1)'_s$, and
neutral-N model $U(1)_N$ charges. \label{E6charge}}
\begin{center}
\begin{tabular}{|c| c| c| c| c|c|c|}
\hline $SO(10)$ & $SU(5)$ & $2 \sqrt{10} Q_{\chi}$ & $2 \sqrt{6}
Q_{\psi}$ & $2 \sqrt{15} Q_{\eta}$ & $2 \sqrt{15} Q_s$ & $2 \sqrt{10} Q_N$\\
\hline
16   &   $10~ (u,d,{\overline u}, {\overline e} )$ & $-$1 & 1  & $-$2 & $-{1/2}$ &1\\
            &   ${\overline 5}~ ( \overline d, \nu ,e)$  & 3  & 1 & 1 & 4 & 2    \\
            &   $1 \overline N$             & $-5$ & 1  & $-5$ & $-5$     & 0    \\
\hline
       10   &   $5~(D,H^{\prime}_u)$    & 2  & $-2$ & 4 & 1    & $-$2      \\
            &   ${\overline 5} ~(\overline D, H^{\prime}_d)$ & $-2$ &$-2$ & 1 & $-{7/2}$ & $-$3 \\
\hline
       1    &   $1~ S_L$                  &  0 & 4 & $-5$ & $5/2$& 5 \\
\hline
\end{tabular}
\end{center}
\end{table}

Besides the general discussion of $E_6$ models, we are interested
in a secluded sector model proposed in ~\cite{ZZ,EWPT},
with  $\theta = \tan^{-1}( \sqrt{15}/9)\sim 0.13 \pi$.
 This model makes use
of three {\bf 27}'s of $E_6$ and some particle pairs from
${\bf 27}+\overline{{\bf 27}}$. It is anomaly free, can solve the MSSM $\mu$
problem, can give a natural explanation of the $Z-Z'$ mass
hierarchy, and allows an enhanced possibility for electroweak baryogenesis.

We focus on the direct search for $Z'$ at hadron colliders such as
the Tevatron and LHC by their decay into $e^-e^+$ and $\mu^-
\mu^+$. We discuss the effect of exotic particles as well as
supersymmetric partners on the discovery limits.
Our concern is not so much the precise discovery limits themselves.
These will be obtained much better by the experimenters
in their own analysis. Rather, we want to quantitatively show the sensitivity of the
discovery limit to the exotic particle spectrum.

The paper is organized as follows. In the next two Sections, we give
the general discussion of $Z'$ physics and the necessary formulas
for the decay width and couplings for mass eigenstates.  In Section \ref{limits} we
study the branching ratios  and
the total decay widths for different $\theta$, and
consider  how the discovery
limits will be affected by the masses of the non-SM particles. In
Section \ref{secluded} we review the secluded sector $E_6$ model and
calculate the discovery limits for $Z'$ only decaying into SM
particles and for decays including the exotics and supersymmetric
partners.

\section{General Discussion of the $Z'$ Decay Width}
\label{general}

The Born cross section $\sigma^f (pp(p\bar{p})\rightarrow
(\gamma,Z,Z')X\rightarrow f \bar{f} X)$ is~\cite{reviews}
\begin{eqnarray}
\sigma^f &=& \Sigma_q \int_0^1 dx_1 \int_0^1 dx_2 \sigma(sx_1
x_2;q \bar{q}\rightarrow f \bar{f})G_A^q(x_1,x_2,M^2_{Z'})
\theta(x_1 x_2 s-M_{\Sigma}^2)
\end{eqnarray}
where $M_{\Sigma}$ is the sum of the masses of the final
particles, $x_{1,2}=\sqrt{{Q^2 \over s}} \exp(+y)$ and $y$ is the
rapidity. The function $G_A^q(x_1,x_2,M^2_{Z'})$ depends on the
structure functions of the quarks. In an approximation adequate for
our purposes,
$\sigma^f$ is given by~\cite{Zlimit,delAguila:1989be}
\begin{eqnarray}
\sigma^f_T \equiv {N_Z' \over L} &=& {1 \over s} c_{Z'} C \exp(-A
{M_{Z'} \over \sqrt{s}}) \label{cros}
\end{eqnarray}
where $C$=600(300) and $A$=32 (20) for $pp$($p\bar{p}$)
collisions. $s$ is the center of mass energy square of the
collision, $N_Z'$ is the number of events, $L$ is the luminosity,
and the subscript $T$ means it is a tree level result. QCD
corrections~($K$ factors) increase the lowest order cross section
by 20-30 percent. We take $K \sim 1.3$ in our numerical
calculation~\cite{leike}. From (\ref{cros}), we see that the predicted cross
section falls exponentially as a function of $M_{Z'}$. The details
of the $Z'$ model are collected in a quantity $c_Z'$ that depends
on $M_{Z'}$, the $Z'$ couplings, and the masses of the particle
the $Z'$ can decay into:
\begin{eqnarray}
c_{Z'} &=& {4 \pi^2 \over 3} {\Gamma_{Z'} \over M_{Z'}} Br_2^f
[Br_2^u+{1 \over C_{ud}} Br_2^d], \label{cz}
\end{eqnarray}
where $C_{ud}=2 (25)$, $\Gamma_{Z'}$ is the total $Z'$ width, and
$Br_2^f$ is the branching ratio into $f\bar{f}$.

 In the limit
that the fermion masses are small compared with $M_{Z'}$, the $Z'$
decay width into fermions is,
\begin{eqnarray}
\Gamma_{Z'\rightarrow  f \bar{f}} &=& {g'^2 M_{Z'} \over {24
\pi}}(Q_l^2+Q_r^2) \label{zf0}
\end{eqnarray}
where $Q_l,Q_r$ are the $U(1)'$ charges for the left (right) chiral
fermions. From (\ref{cz}), the
dependence on $M_{Z'}$ cancels out in this limit, and $c_Z'$ is a constant
depending only on the particle charge assignments. Then
(\ref{cros}) can be inverted to obtain the \zpr\ discovery limit,
\begin{eqnarray}
M_{Z'}^{lim} \sim {\sqrt{s} \over A} \ln \left({L \over s} {c_{Z'}
C \over N_{Z'}}\right) \sim \sqrt{s} \times 0.386 (0.583) +{1
\over 32 (20)} \ln \left({L \cdot fb \over N_{Z'} \cdot s/TeV^2}
\cdot 1000 ~c_{Z'}\right) \label{Mzlim}
\end{eqnarray}
For $M_{Z'}<M_{Z'}^{lim}$, more than $N_{Z'}$ events are expected.
If the $Z'$ decays only into the SM fermions,  (\ref{Mzlim}) gives
a good estimate. The result is changed only a small amount for the models
and mass ranges we
are considering if we include the effect of the top quark mass of
175 $GeV$.

However, to study the effects of light exotic particle and/or
light supersymmetric partners on the $Z'$ discovery limits, we
need to deal with particles with masses comparable to the $Z'$
mass. In that case, the $Z'$ decay width will depend on the
particle masses, and  $c_{z'}$ is not a constant, implying the
need for a numerical study. The width for $Z'$ decays into boson
pairs with nonzero masses is~\cite{Ztoboson}\cite{neum}
\begin{eqnarray}
\Gamma_{Z\rightarrow  s_1 s_2^*} &=& {g'^2 f_{z s_1 s_2^*}^2
M_{Z'} \over {48 \pi}} \left(1+{m_1^4 \over M_{Z'}^4}+{m_2^4 \over
M_{Z'}^4}-2{m_1^2 \over M_{Z'}^2}-2 {m_2^2 \over M_{Z'}^2}-2
{m_1^2 m_2^2\over M_{Z'}^4}\right)^{3\over 2} \label{Zbb}
\end{eqnarray}
where $f_{z s_1 s_2^*}$ is the $Z'$ couplings to $s_1,s_2^*$. It
is a product of \upr\ charges and a matrix connecting the mass
eigenstates and weak eigenstates and will be given in the next
section. The width for $Z'$ decays into fermions is more
complicated because the formulae for $Z'$ decay into Majorana
spinors~(for example, neutralinos) and Dirac spinors ~(quarks and
leptons) are different. The decay width for $Z'$ decays into
Majorana spinors is ~\cite{neum}
\begin{eqnarray}
\Gamma_{Z'\rightarrow  f_i f_j} &=& {g'^2 M_{Z'} \over {24
\pi}}\left(1+{m_i^4 \over M_{Z'}^4}+{m_j^4 \over M_{Z'}^4}-2{m_i^2
\over M_{Z'}^2}-2 {m_j^2 \over M_{Z'}^2}-2 {m_i^2 m_j^2\over
M_{Z'}^4}\right)^{1\over 2} \nonumber\\&&
\left[\left(1-{m_i^2+m_j^2\over 2 M_{Z'}^2}-{(m_i^2-m_j^2)^2\over
2 M_{Z'}^4}\right)(C_{l(i,j)}^2+C_{r(i,j)}^2)\right. \nonumber\\&&
\left. +3 {m_1 m_2 \over M_{Z'}^2}(C_{l(i,j)}
C_{r(i,j)}^*+C_{r(i,j)} C_{l(i,j)}^*)\right] \frac{1}{1+\delta_{ij}}
\label{Zff}
\end{eqnarray}
where $C_l$ and $C_r$ are the product of \upr\ charges and the
matrix connecting the mass and weak eigenstates. They will be
given in the next section. $m_{1,2}$ are the masses of the two
fermions. From the above, we can deduce the decay width for $Z'$
decay into Dirac fermion pairs,
\begin{eqnarray}
\Gamma_{Z'\rightarrow  f \bar{f}} &=& {g'^2 M_{Z'} \over {24
\pi}}\left(1-4{m^2 \over M_{Z'}^2}\right)^{1\over
2}\left[\left(1-{m^2\over M_{Z'}^2}\right)(C_l^2+C_r^2) \right. \nonumber \\
&+& \left. 3 {m^2
\over M_{Z'}^2}(C_l C_r^*+C_r C_l^*)\right] \label{ZffD}
\end{eqnarray}
We now consider the couplings between mass eigenstates.

\section{Couplings Between Mass Eigenstates}
\label{eigenstates}

If the Higgs fields have nonzero $U(1)'$ charge, there will be
mixing between the two neutral gauge bosons. The covariant
derivative appearing in the Lagrangian of a supersymmetric $U(1)'$
model (we only include neutral gauge bosons) is\footnote{We neglect the
possibility of kinetic mixing~\cite{kinetic}.}
\begin{eqnarray}
D_{\mu} &=& \partial_{\mu}-i{g_1 g_2 \over \sqrt{g_1^2+g_2^2}}
A_{\mu} (T^3+Y)-i {1 \over \sqrt{g_1^2+g_2^2}} Z_{\mu} (g_2^2 T^3
-g_1^2 Y)-i g' Q' Z'_{\mu}
\end{eqnarray}
After the symmetry breaking, $H_{u,d}$ acquire nonzero VEVs
$v_{1,2}$ and some SM singlets $S_i$ that are charged under $U(1)'$
will acquire nonzero VEVs to account for the $Z-Z'$ mass
hierarchy. The $Z-Z'$ mass squared matrix is
\begin{eqnarray}
 M_{Z Z'}^2  &=&  \left( \begin{array}{c c}
M_1^2 & \delta M^2 \\
\delta M^2 & M_2^2
  \end{array} \right)
\end{eqnarray}
where
\begin{eqnarray}
\delta M^2 &=& \sqrt{g_1^2+g_2^2} g' (Q'_{h_2} v_2^2-Q'_{h_1}
v_1^2)
\end{eqnarray}
\begin{eqnarray}
M_1^2 &=& {(g_1^2+g_2^2) (v_1^2+v_2^2) \over 2}
\end{eqnarray}
\begin{eqnarray}
M_2^2 &=& 2 g'^2 (Q'^2_{h_1} v_1^2+Q'^2_{h_2} v_2^2+Q'^2_{S_i}
S_i^2)
\end{eqnarray}
The mass eigenstates $Z_1,Z_2$ are related to $Z$ and $Z'$ by
\begin{eqnarray}
 \left( \begin{array}{c} Z \\ Z' \end{array} \right)
 &=&  \left( \begin{array}{c c}
\cos{\theta_z} & \sin{\theta_z} \\
- \sin{\theta_z} & \cos{\theta_z}  \end{array} \right)
\left(\begin{array}{c} Z_1 \\ Z_2 \end{array} \right)
\end{eqnarray}
where $\theta_z$ is the $Z-Z'$ mixing angle, given by
\begin{eqnarray}
\tan{2 \theta_z} &=& {2 \delta M^2 \over M_2^2 - M_1^2}
\end{eqnarray}
The covariant derivative in terms of $Z_1$ and $Z_2$ is
\begin{eqnarray}
D_{\mu} &=& \partial_{\mu}-i{g_1 g_2 \over \sqrt{g_1^2+g_2^2}}
A_{\mu} (T^3+Y)-i \left({\cos{\theta_z} \over
\sqrt{g_1^2+g_2^2}}(g_2^2 T^3 -g_1^2 Y)-g' Q'
\sin{\theta_z}\right) Z_{1 \mu} \nonumber \\&& - i \left(g'Q'
\cos{\theta_z}+{\sin{\theta_z} \over \sqrt{g_1^2+g_2^2}}(g_2^2 T^3
-g_1^2 Y)\right) Z'_{2 \mu}
\end{eqnarray}
The mixing angle, $\theta_z$, is very small by the LEP and SLD
$Z$-pole data and other precise constraints~\cite{indirect1}-\cite{cheung},
so we will neglect it
in the following\footnote{Because we have neglected the mixing
angle, we don't consider the decays $Z' \rightarrow W^+ W^-$ and
$Z' \rightarrow Z+boson$, since the amplitudes are proportional to
the mixing angle~\cite{leike,DEE6,Ztoboson,gunion}.}.

We also need the $Z'$ couplings to the mass eigenstates. For the
standard model and other Dirac fermions, the $C_l$ and $C_r$ in
(\ref{ZffD}) are just the \upr\ charges for the left-handed and
right-handed components. For squarks and sleptons, the Lagrangian
is
\begin{eqnarray}
L &=& -i g' Q'_L Z'_{\mu} \phi_L \partial^{\mu} \phi^*_L-i g' Q'_R
Z'_{\mu} \phi_R \partial^{\mu} \phi^*_R \label{zss}
\end{eqnarray}
For simplicity, we neglect family mixing here. Let
$\phi_1,\phi_2$ be the mass eigenstates and $\theta_s$ the mixing
angle between the left and right fields.
\begin{eqnarray}
 \left( \begin{array}{c} \phi_L \\ \phi_R \end{array} \right)
 &=&  \left( \begin{array}{c c}
\cos{\theta_s} & \sin{\theta_s} \\
- \sin{\theta_s} & \cos{\theta_s}  \end{array} \right)
\left(\begin{array}{c} \phi_1 \\ \phi_2 \end{array} \right)
\end{eqnarray}
In terms of the mass eigenstates,
\begin{eqnarray}
L &=& -i g' Z'_{\mu} [(Q'_L \cos^2{\theta_s}+Q'_R
\sin^2{\theta_s}) \phi_1 \partial^{\mu} \phi^*_1+(Q'_L
\sin^2{\theta_s} + Q'_R \cos^2{\theta_s})  \phi_2 \partial^{\mu}
\phi^*_2 \nonumber \\&& +(Q'_L-Q'_R) \sin{\theta_s} \cos{\theta_s}
(\phi_1 \partial^{\mu} \phi^*_2+\phi_2 \partial^{\mu} \phi^*_1) ]
\label{zhh}
\end{eqnarray}
from which we can read off the couplings,
\begin{eqnarray}
f_{z11^*} &=& (Q'_L \cos^2{\theta_s}+Q'_R \sin^2{\theta_s})
\nonumber\\ f_{z22^*} &=& (Q'_L \sin^2{\theta_s} + Q'_R
\cos^2{\theta_s}) \nonumber\\ f_{z12^*}=f_{z21^*} &=& (Q'_L-Q'_R)
\sin{\theta_s} \cos{\theta_s}.
\end{eqnarray}
It is straightforward to generalize these formulae to include
family mixing. In the massless limit, adding all of the possible
decay channels, the decay width is,
\begin{eqnarray}
\Gamma_{Z'\rightarrow  b b^*} &=& {g'^2 M_{Z'} \over {48
\pi}}(Q_l^2+Q_r^2) \label{zb0}
\end{eqnarray}
There is a simple relation between the decay width into the standard
model fermions and their supersymmetric partners in one chiral
supermultiplet in the massless limit, i.e.,  $\Gamma_{Z'\rightarrow b
b^*}={1 \over 2 } \Gamma_{Z'\rightarrow f \bar{f}}$.

The couplings between mass eigenstates for other fields can be
obtained parallel to the above discussion.  There are some
subtleties in the extended Higgs and neutralino sectors. Let
$H_i$, $A_i$ be the mass eigenstates for CP even and CP odd Higgs
fields and $H_{wi}$, $A_{wi}$ the weak eigenstates, with
\begin{eqnarray}
H_i &=& U_{ij} H_{wi}  ~,~   A_i=V_{ij} A_{wj} ~,
\end{eqnarray}
where $U_{ij}$ and $V_{ij}$ are unitary matrices. Then
\begin{eqnarray}
L &=& i g' H_j U^{-1}_{ji} Q'_i V^{-1}_{ik} \partial A_k
\end{eqnarray}
The mass matrix can be derived directly from the superpotential.
For the $E_6$ model with a secluded sector, it is given in the
appendix of ~\cite{EWPT}.

For the neutralino sector, since they are Majorana spinors, we
need to pay more attention to the details of their couplings~(For the
chargino sector, we will obtain similar formulae; see ~\cite{neum}
for details.). Let $N_i$ be the weak eigenstates and $\chi_i$ the
4-component Majorana spinor mass eigenstates, and
\begin{eqnarray}
\psi_i \equiv  \left( \begin{array}{c} N_i \\ \overline{N}_i
\end{array} \right) &=& N_{ij} \chi_{wj}
\end{eqnarray}
Then,
\begin{eqnarray}
L &=& -i g' Q_i' Z^{\mu} \bar{N_i} \bar{\sigma}^{\mu} N_i
\nonumber \\&& ={i \over 2} g' Z^{\mu} \bar{\chi_i} \gamma^{\mu}
(C_{l(i,j)} P_L+C_{r(i,j)} P_R ) \chi_j
\end{eqnarray}
where $C_{l(i,j)}=Q'_k N_{ik} N_{jk}^* ~,~
C_{r(i,j)}=-{C^*_{l(i,j)}}$, and $N_{ij}$ is the matrix that
diagonalizes the neutralino mass matrix. $C_{l(i,j)}$ and
$C_{r(i,j)}$ are the couplings appearing in (\ref{Zff}). One can
obtain the matrix from~\cite{EWPT} for the $E_6$ model with a
secluded sector. In that model, since there are 4 Higgs singlet
fields, the mass matrices of the Higgs and neutralinos are $6
\times 6$ and $9 \times 9$, respectively.

\section{$Z'$ Decay in Different Supersymmetric $E_6$ models }
\label{limits}
Including more particles will enlarge the total decay width for
the $Z'$ and reduce the branching ratio to quark and lepton pairs,
reducing the discovery limit. As shown in
 (\ref{Zbb}) and  (\ref{Zff}), the partial decay width depends on
the particle mass as well as the couplings to the $Z'$. If the sum
of the particle masses is larger than the $Z'$ mass, the decay
will not be kinematically allowed. Also, the $Z'$ couplings will be
affected by the mass matrix through mixing, as can be read off
from (\ref{zss}) and  (\ref{zhh}).

\begin{figure}
\begin{center}
\caption{The total decay width for $E_6$ models as a function of
the angle $\theta$. The solid (dotted) lines are respectively for
the case in which all the exotic and supersymmetric partners are
massless and for decays into standard model fermions only. $m_t$
is neglected.}
\vspace{0.4cm}
\includegraphics[height=3.6in]{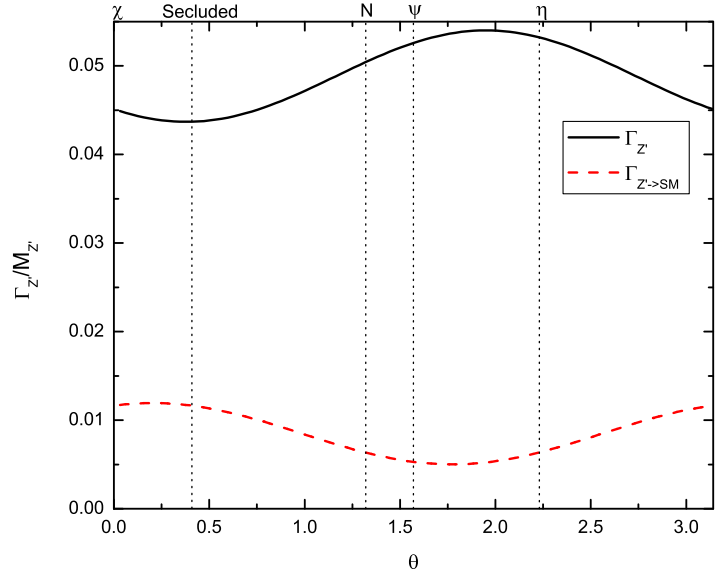}
\label{DEWI}
\end{center}
\end{figure}

\begin{figure}
\caption{The branching ratios into SM particles for $E_6$ models
as a function of  $\theta$. The first graph is for the case that
$Z'$ only decays into SM fermions, neglecting $m_t$. The second assumes that $Z'$
can decay into all of the particles and that every particle is
massless.} \vspace{0.4cm}
\includegraphics[height=2.5in]{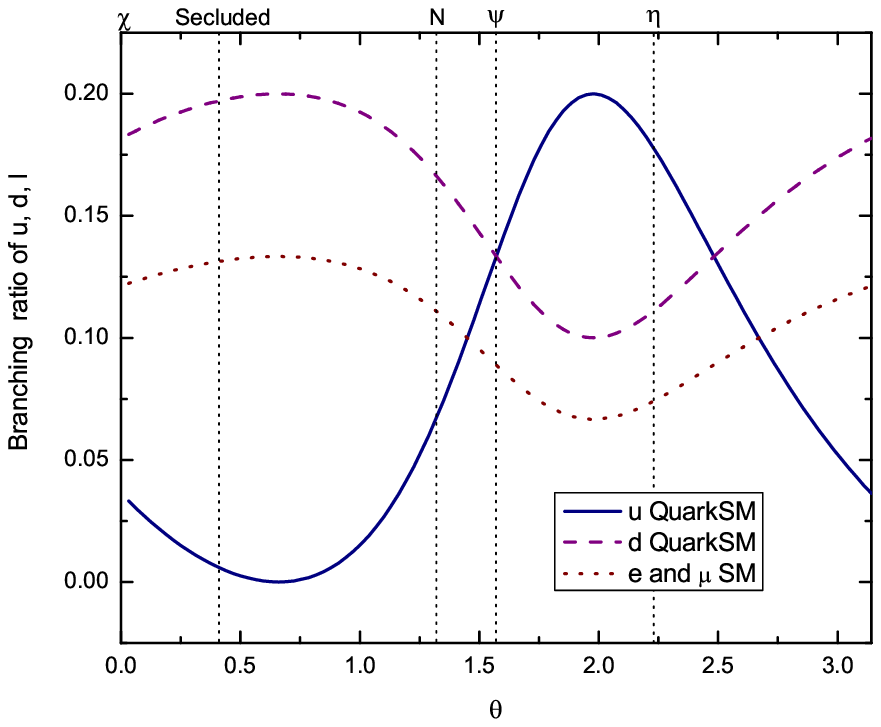}
\includegraphics[height=2.5in]{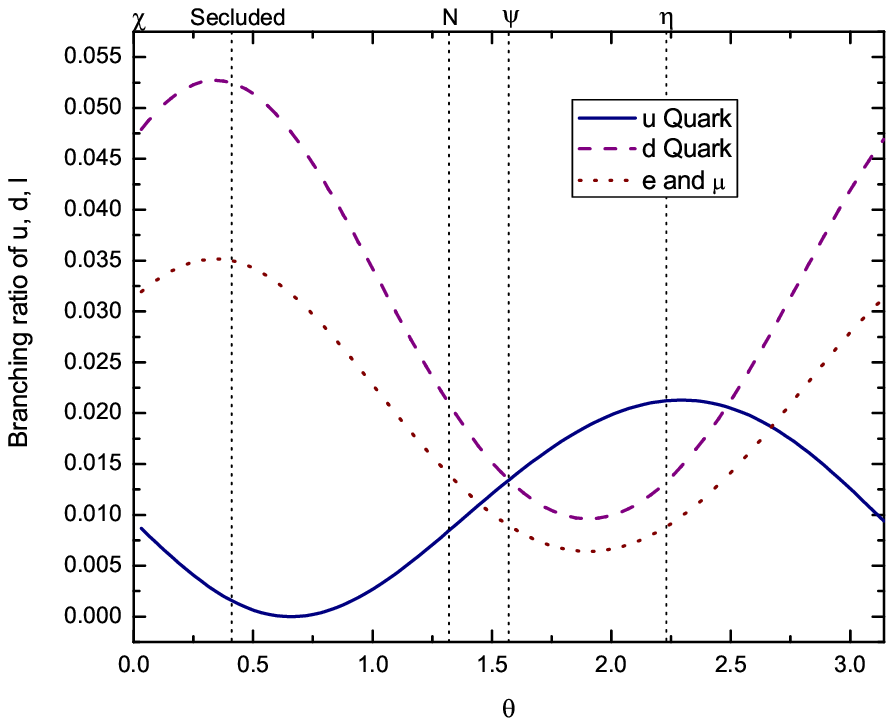}
\label{BrEQ}
\end{figure}

For a general $E_6$ model, the $U(1)'$ is a linear combination of
$U(1)_{\chi}$ and $U(1)_{\psi}$, as in  (\ref{E6MIX}). The angle
$\theta$ will affect the particle charges and the total decay
width (Fig. \ref{DEWI}), and branching ratios (Fig. \ref{BrEQ}). The
dashed line in Fig. \ref{DEWI} is the total width for $Z'$ to
decay only into SM fermions as a function of $\theta$. From
(\ref{zf0}) and (\ref{zb0}), the decay width is proportional to
the sum of charge squares of the quarks and leptons. The solid
line is a limiting case in which all of the particles are
massless. In that case, the decay width will involve the sum of
charge squares of all of the particles. If the particle content
only included three copies of {\bf 27}, the charge square sum would be a
constant, independent of the $\theta$. We include an additional
$H_u,\overline{H}_u$ from the ${\bf 27}+\overline{{\bf 27}}$, as suggested by
the gauge coupling unification~\cite{ZZ}, and three singlet pairs
from the ${\bf 27}+\overline{{\bf 27}}$ as in the $E_6$ model with a secluded
sector~\cite{EWPT}. The charge square sum then depends on
$\theta$, due to the charges of $H_u,\overline{H}_u$ and the
singlets.

In Fig. \ref{BrEQ}, the branching ratios of up quarks, down quarks
and the sum of electrons plus muons are shown as a function of $\theta$.
These affect the discovery limit as suggested by (\ref{cz}) and
(\ref{Mzlim}). The first graph is the branching ratio in the case
that the $Z'$ only decays into SM fermions. The second is for the
case that the $Z'$ can decay into all of the particles and that
every particle is massless.

\begin{figure}
\caption{Left: the branching ratios into Higgs and exotic particles
and their superpartners  as
a function of $\theta$. $D$, $H$, and $S$ are the
quark-like, Higgs-like, and singlet particles. Family and color
degeneracy have been included. $H_u$ and $\overline{H}_u$ from
${\bf 27}+\overline{{\bf 27}}$ have been included in the $H$, and $3$ pairs of
singlets from the  ${\bf 27}+\overline{{\bf 27}}$ have been included in the
$S$. Right: the branching ratios into SM particles and their
superpartners. Family and color degeneracy have been included.
Lepton includes charged leptons and left-handed neutrinos. }
\vspace{0.4cm}
\includegraphics[height=2.5 in]{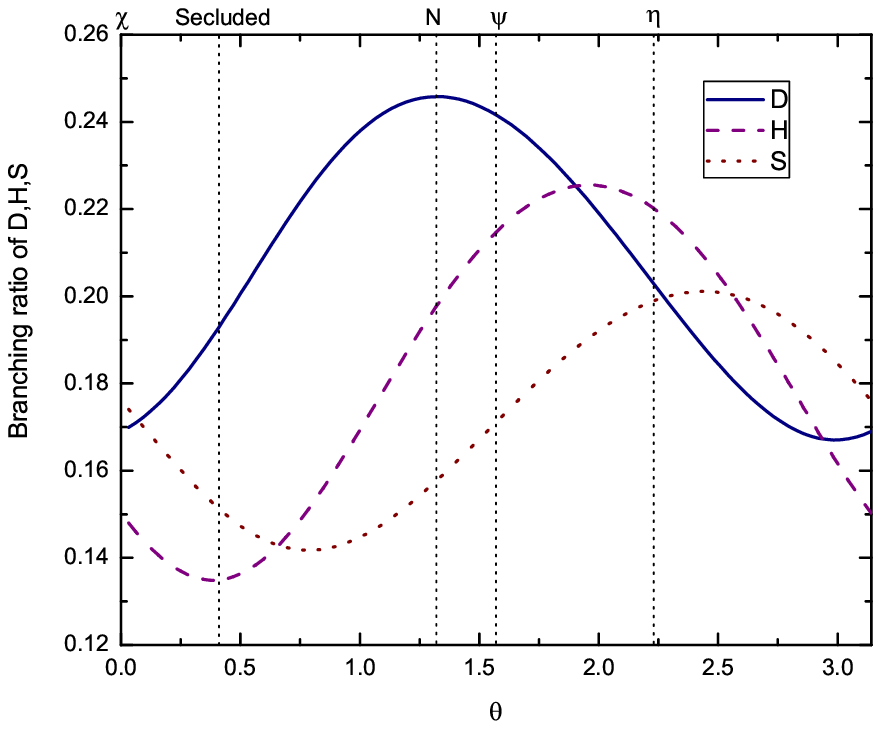}
\includegraphics[height=2.5 in]{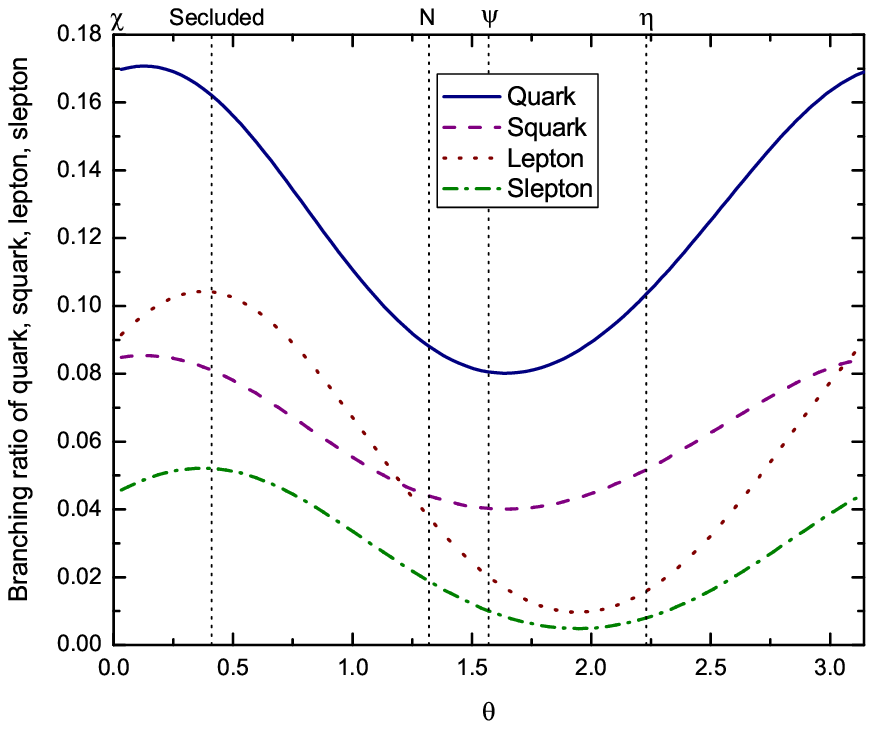}
\label{BrEX}
\end{figure}

Including exotic particles and the superpartners of SM particles
enlarges the total decay width and reduces the branching ratio for
quarks and leptons significantly, i.e., by factors of 5 to 10. The
exotic branching ratios are displayed in Fig. \ref{BrEX}, assuming
that all of the particles are massless. We classify the exotic
particles as quark-like, Higgs-like (including the Higgs), and SM
singlets, and consider the combined contributions of $Z'$ decaying
into the exotics and their superpartners. In the second graph in
Fig. \ref{BrEX}, we display the branching ratios of  the
 superpartners of the quarks (summing the three families),
the superpartners of the leptons (charged
leptons and left-handed neutrinos), as
well as those of the quarks and the leptons.

\begin{figure}
\caption{The $Z'$ discovery limit (for 10 dilepton events) for the
$E_6$ model at the Tevatron ($\sqrt{s}$=1.96 TeV and $L=1,3$
$fb^{-1}$). Lines with the label min are the limiting case in
which all of the particle are massless; lines without are the
cases in which the $Z'$ decays only into SM fermions, and the
physical top quark mass is included. The intersection point
between the experimental line and the theoretical lines are the
discovery limits. The particle content is $3 \times {\bf 27}$;
$H_u, \overline{H}_u$ from ${\bf 27}+\overline{{\bf 27}}$; and 3
pairs of singlets. The sequence of the curves is same as the
sequence in the legend. The lines $N_{min}$ and $\psi_{min}$ are almost
on top of each other.} \vspace{0.4cm}
\begin{center}
\includegraphics[height=4.0 in]{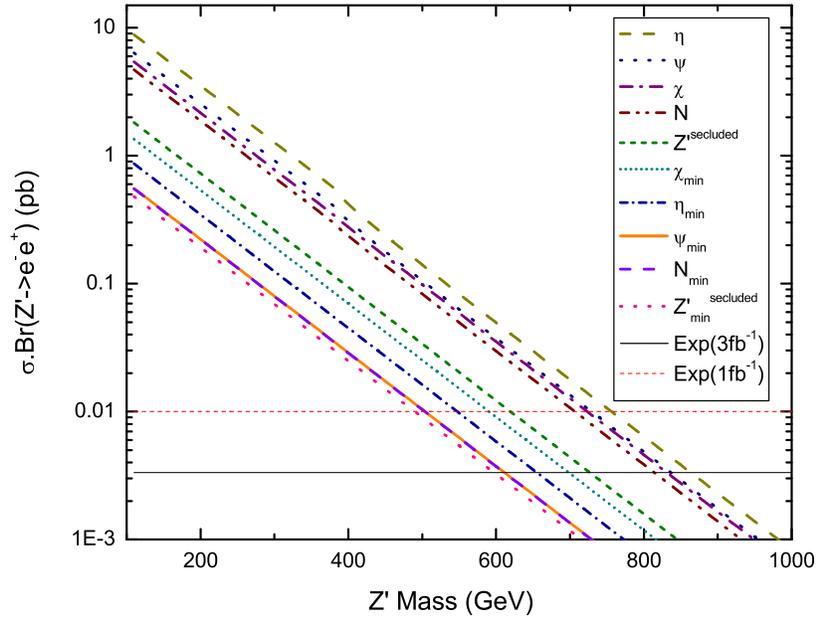}
\label{Teva}
\end{center}
\end{figure}

\begin{figure}
\caption{The $Z'$ discovery limit (GeV) for the $E_6$ model at the
LHC ($\sqrt{s}$=14 TeV and $L=100,300$ $fb^{-1}$).The sequence of
the curves is same as the sequence in the legend. The lines $N$ and $\eta$ and the lines  $N_{min}$ and
$\eta_{min}$ are almost on top of each other.} \vspace{0.4cm}
\begin{center}
\includegraphics[height=4.0 in]{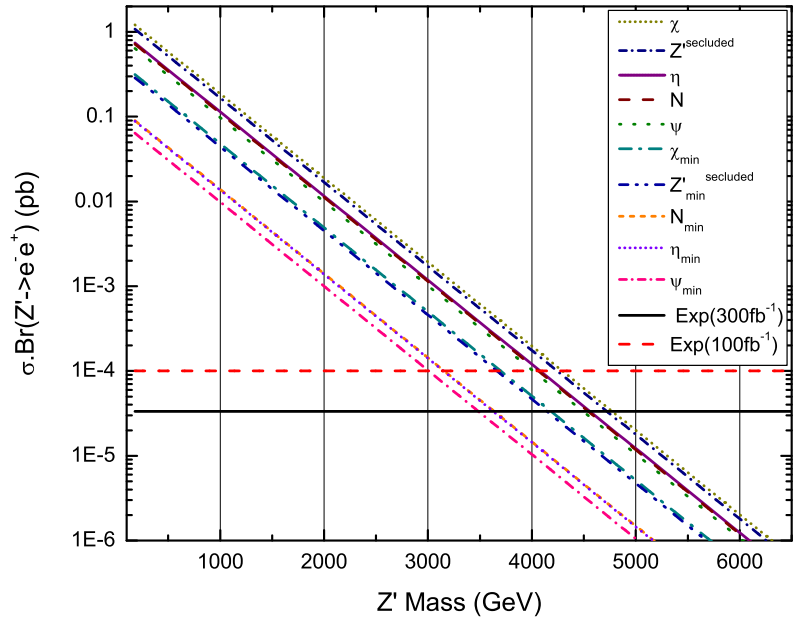}
\label{LHC}
\end{center}
\end{figure}

We now consider the discovery limits for $E_6$ models at the
Tevatron and LHC.
The \upr\ charges for specific $E_6$ models can be found in Table
\ref{E6charge}.
We draw two limiting cases in Fig. \ref{Teva}
for the Tevatron ($\sqrt{s}$=1.96 TeV and $L=1,3$ $fb^{-1}$) and
Fig. \ref{LHC} for the LHC ($\sqrt{s}$=14 TeV and $L=100,300$
$fb^{-1}$). One is just the $Z'$ decay into SM fermions. The other
is a limiting case in which all of the particle are massless.The
particle content is $3 \times {\bf 27}$ and $H_u, \overline{H}_u$ from
${\bf 27}+\overline{{\bf 27}}$, which comes from the gauge unification
requirement. We also include $3$ pairs of singlets for comparison
with the secluded sector $E_6$ model.

We have included the top quark mass effect in the SM figures. We
use (\ref{cros}) as an estimate of the cross section; it is not
accurate for small $M_{Z'}$ but gives a good approximation for the
large $M_{Z'}$ that we are mainly concerned with. The QCD $K$
factor has been included. The upper limit experimental line is
based on a total of 10 dilepton events, i.e., including both
$e^+e^-$ and $\mu^+\mu^-$. This is meant to be a rough
idealization to illustrate the effect of the exotics and
sparticles. Of course, the actual experimental analysis leads to
an experimental line with more complicated structure than the
horizontal line in the figure.

The intersection points between the experimental line and the
theoretical lines are the discovery limits, which are shown in
Table \ref{Teval}. The discovery limits at the Tevatron for the $E_6$ model with a
secluded sector are lower than the other $E_6$
models because the $U(1)'$ charge of the up quark is smaller. For
comparison, we also show the discovery limit for different $E_6$
models with the particle content $3\times {\bf 27} $ and $H_u,
\overline{H}_u$ (but without the extra pairs of singlets)
 in Table \ref{LHCl}. Since the number
of exotic particles is less than that of the previous case, the
discovery limits are increased.

\begin{figure}
\caption{The discovery limits for $E_6$ models as a function of
$\theta$ at the Tevatron and the LHC. The lower two curves are the
discovery limits in the case that $Z'$ decays into all of
particles and all are massless (with $L=1 fb^{-1}$
and $L=3fb^{-1}$ for the Tevatron and $L=100 fb^{-1}$ and
$L=300fb^{-1}$ for the LHC). The upper two curves assume that $Z'$
only decays into SM fermions, with the top quark mass included.}
\vspace{0.4cm}
\includegraphics[height=2.5 in]{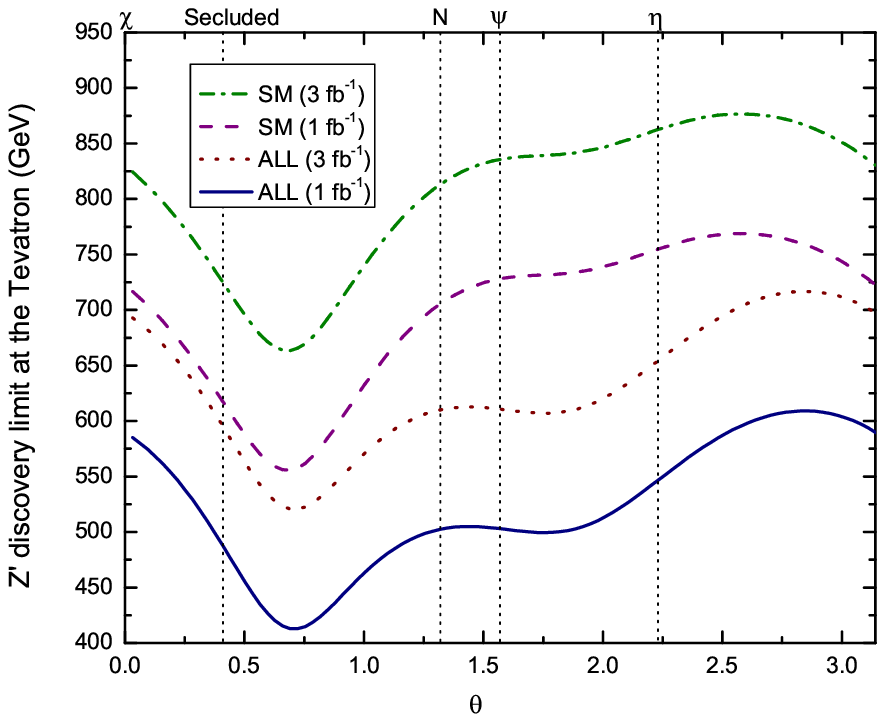}
\includegraphics[height=2.5 in]{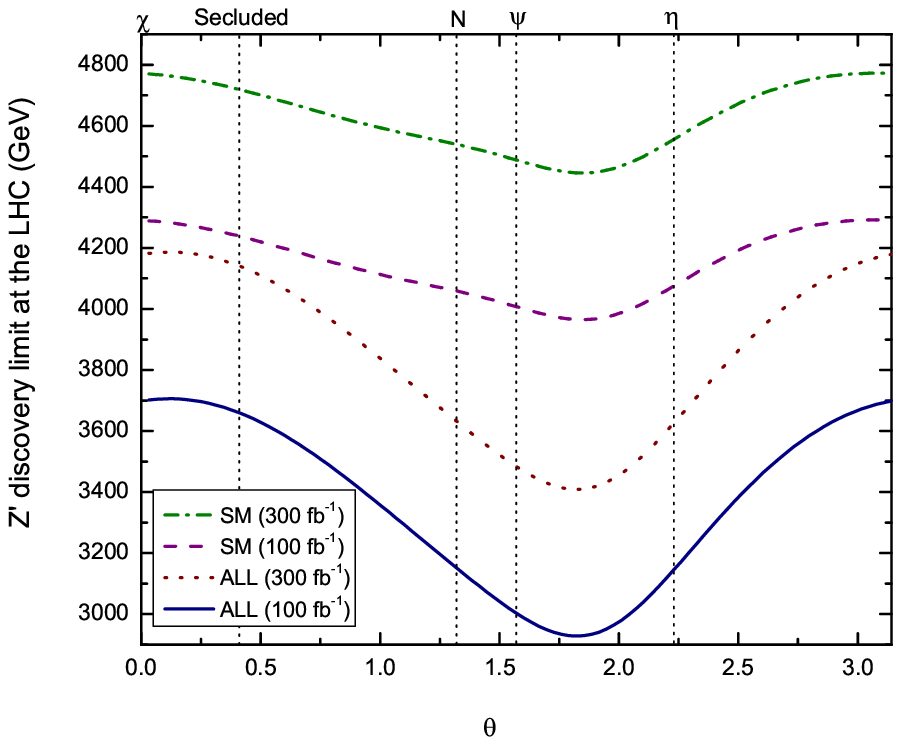}
\label{dis}
\end{figure}

\begin{table}[t]
\caption{ The $Z'$ discovery limits (GeV) (for 10 dilepton events)
for the $E_6$ model at the Tevatron ($\sqrt{s}$=1.96 TeV and
$L=1~(3)$ $fb^{-1}$) and LHC ($\sqrt{s}$=14 TeV and $L=100~(300)$
$fb^{-1}$), corresponding to Fig. \ref{Teva} and Fig. \ref{LHC}.
The particle content is $3 \times {\bf 27}$; $H_u, \overline{H}_u$
from ${\bf 27}+\overline{{\bf 27}}$; and 3 pairs of singlets. \label{Teval}}
\begin{center}
\begin{tabular}{|c| c| c| c| c| c|}
\hline $Tevatron$ & ${Z'}^{secluded}$ & $\chi$ & $\psi$ & $\eta$ & $N$\\
\hline $extreme$ & 490(598)& 591(699)& 505(612)& 548(656) & 503(611)\\
\hline $Z' \to SM$ & 619(727)& 724(832)& 730(837)& 757(864) & 706(814)\\
\hline $LHC$ & ${Z'}^{secluded}$ & $\chi$ & $\psi$ & $\eta$ & $N$\\
\hline $extreme$ & 3665(4147)& 3703(4184)& 3008(3489)& 3150(3631) & 3157(3638)\\
\hline $Z' \to SM$ & 4243(4725)& 4295(4776)& 4012(4493)&4079(4561) & 4063(4544)\\
\hline
\end{tabular}
\end{center}
\end{table}
\begin{table}[t]
\caption{ Same as Table \ref{Teval}, but without the 3 pairs of singlets. \label{LHCl}}
\begin{center}
\begin{tabular}{|c| c| c| c| c| c| }
\hline $Tevatron$ & ${Z'}^{secluded}$ & $\chi$ & $\psi$ & $\eta$ & $N$\\
\hline $extreme$ & 500(610)& 600(710)& 520(638)& 559(676) & 526(634)\\
\hline $Z' \to SM$ & 619(727)& 724(832)& 730(837)& 757(864) & 706(814)\\
\hline $LHC$ & ${Z'}^{secluded}$ & $\chi$ & $\psi$ & $\eta$ & $N$\\
\hline $extreme$ & 3700(4201)& 3718(4230)& 3075(3580)& 3220(3720) & 3260(3741)\\
\hline $Z' \to SM$ & 4243(4725)& 4295(4776)& 4012(4493)&4079(4561) & 4063(4544)\\
\hline
\end{tabular}
\end{center}
\end{table}
From Figs. \ref{Teva} and \ref{LHC}, we see that different \upr\
charge assignments affect the discovery limit significantly. The
discovery limits at the Tevatron and LHC as a function of
$\theta$ are shown in Fig. \ref{dis}. The $\theta$
dependences are quite different.
This is  mainly because  the contribution to the $Z'$ production from
the $u$ quark dominates for the Tevatron,
while  the $u$ and $d$
quark contributions are more comparable for the LHC.

The above analysis assumes that all of the particles are massless
for the limiting case. We also consider varying the masses of the
non-SM particles. As seen in (\ref{Zbb}) and (\ref{Zff}), the mass
enters the discovery limit through the decay width. Since the
$c_{Z'}$ in (\ref{cz}) is no longer independent of $M_{Z'}$, we can't
convert it analytically to obtain the discovery limit as in
(\ref{Mzlim}), but have to study it numerically.

We classify the non-SM particles as the superpartners of standard
model quarks and leptons, the Higgs-like particles, the quark-like
exotics, and singlets. Both fermions and bosons are included in
the last three classes. We take the masses of all the particles and
their superpartners in each class to be degenerate for simplicity,
except that squarks and sleptons have different masses from the
quarks and leptons. We also don't distinguish between the ordinary
and exotic Higgs fields.

\begin{figure}
\caption{The $Z'$ discovery limit for the $\chi$ model at the
Tevatron ($\sqrt{s}$=1.96 TeV and $L$=1,~3 $fb^{-1}$, left graph)
and LHC ($\sqrt{s}$=14 TeV and $L$=100,~300 $fb^{-1}$, right
graph) as a function of the mass of the various non-SM particles.
The lowest curve, labelled {\em Total}, assumes that all of the
particles have a common mass and the others assume only decays
only into the SM fermions and one class of new particle. The
sequence of the curves is same as the sequence in the legend.}
\vspace{0.4cm}
\includegraphics[height=2.5in]{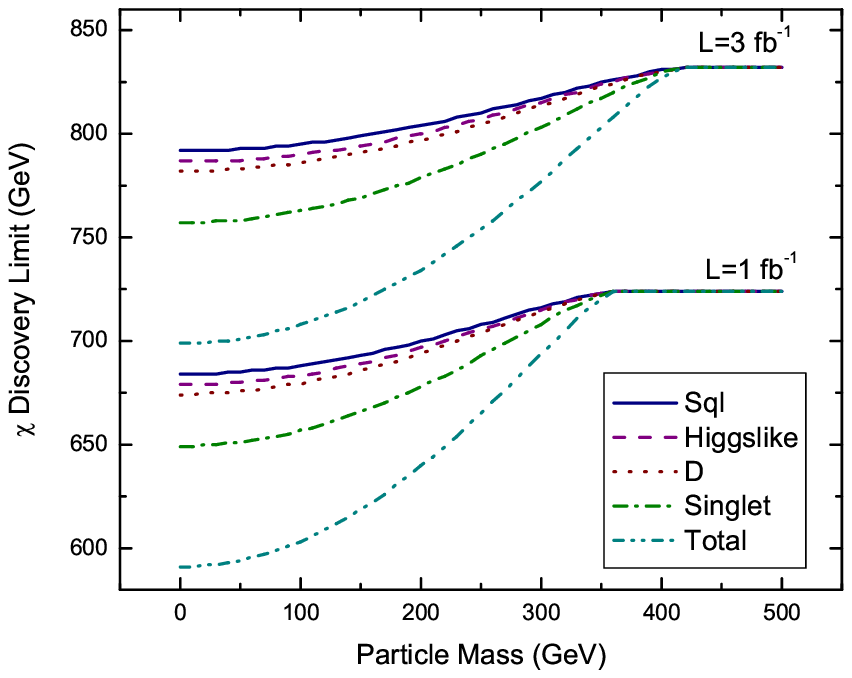}
\includegraphics[height=2.5in]{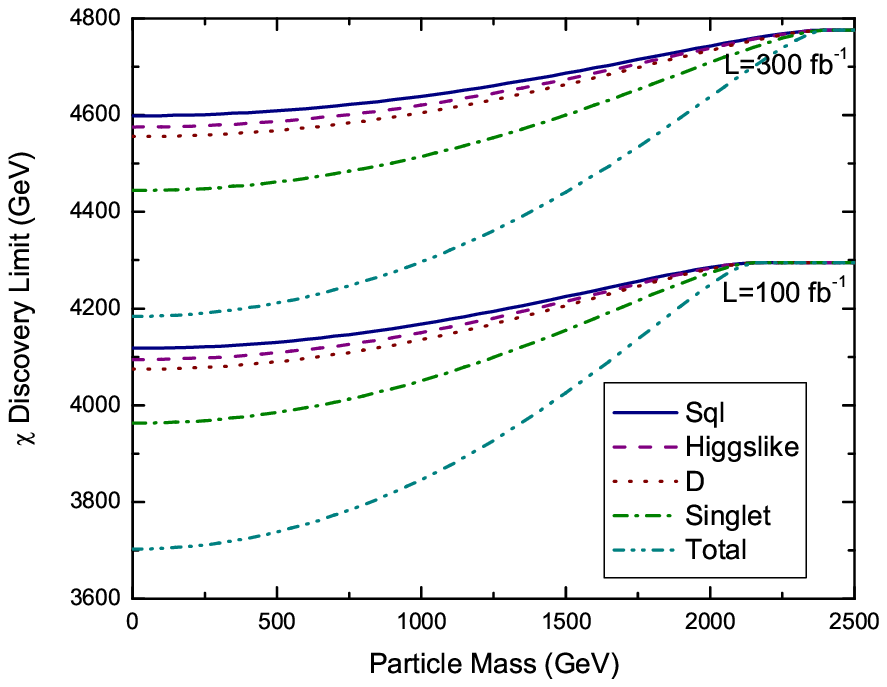}
\label{Echi}
\end{figure}

\begin{figure}
\caption{The $Z'$ discover limit for the $\psi$ model at the
Tevatron and LHC. The sequence of the curves is same as the
sequence in the legend.} \vspace{0.4cm}
\includegraphics[height=2.5 in]{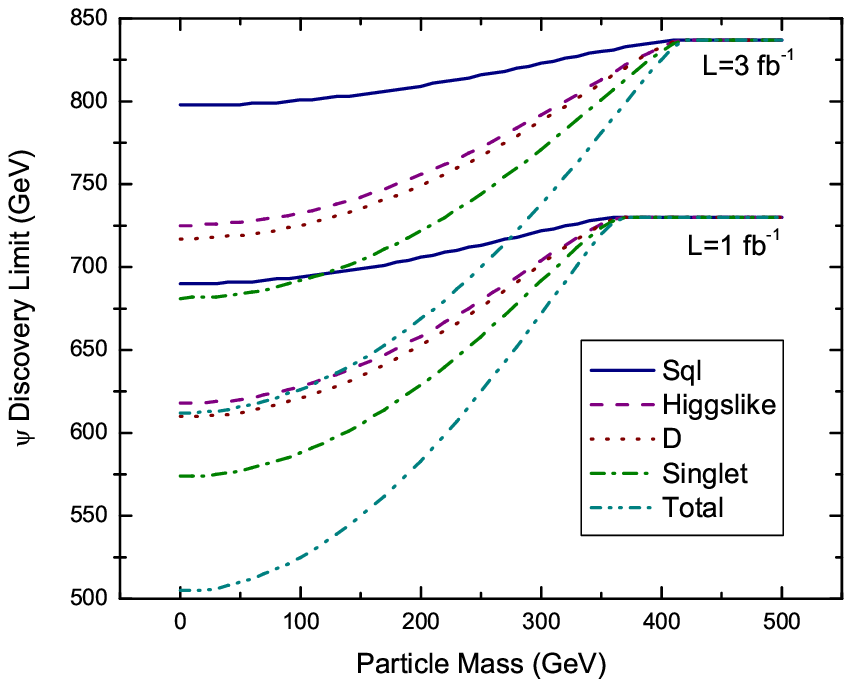}
\includegraphics[height=2.5 in]{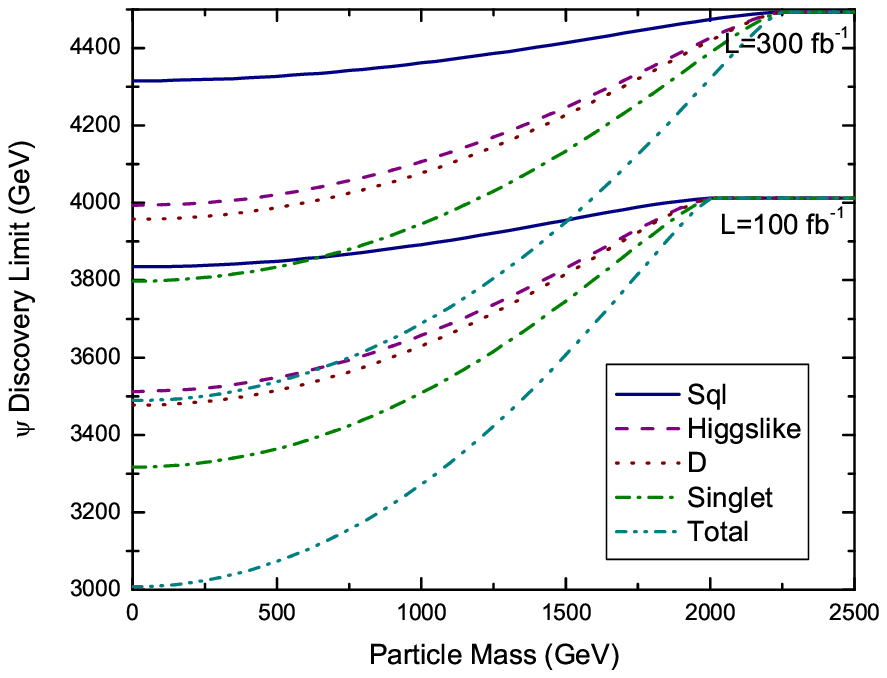}
\label{Epsi}
\end{figure}

\begin{figure}
\caption{The $Z'$ discover limit for the $\eta$ model at the
Tevatron and LHC. The sequence of the curves is same as the
sequence in the legend.} \vspace{0.4cm}
\includegraphics[height=2.5 in]{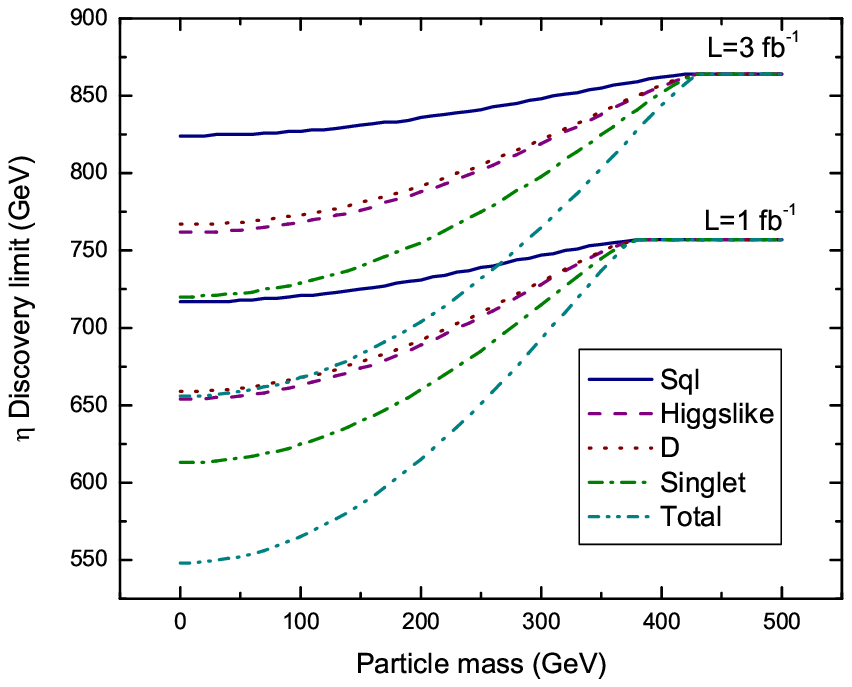}
\includegraphics[height=2.5 in]{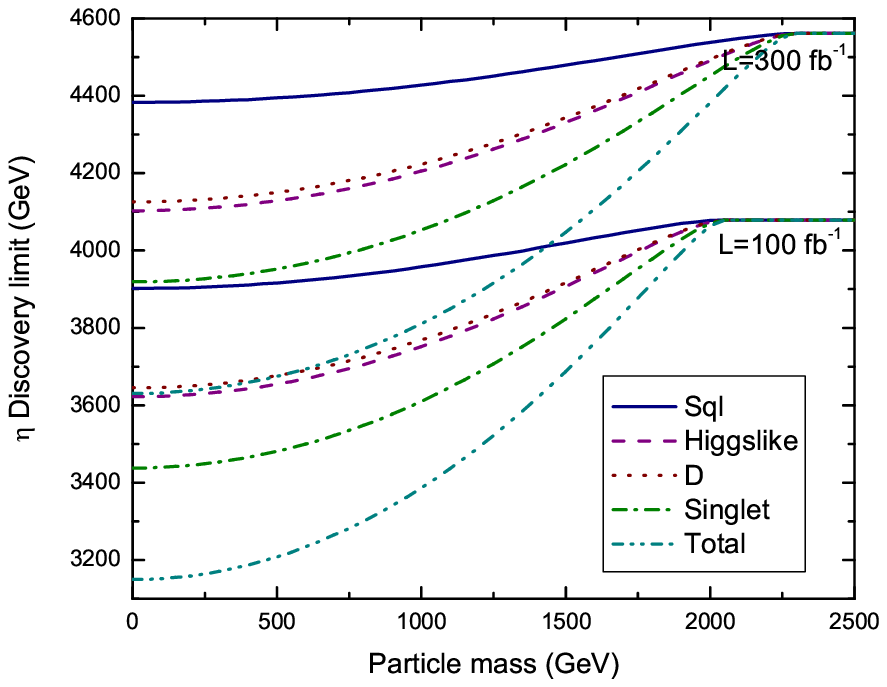}
\label{Eeta}
\end{figure}

\begin{figure}
\caption{The $Z'$ discover limit for the $N$ model at the Tevatron
and LHC. The sequence of the curves is same as the sequence in the
legend.} \vspace{0.4cm}
\includegraphics[height=2.5 in]{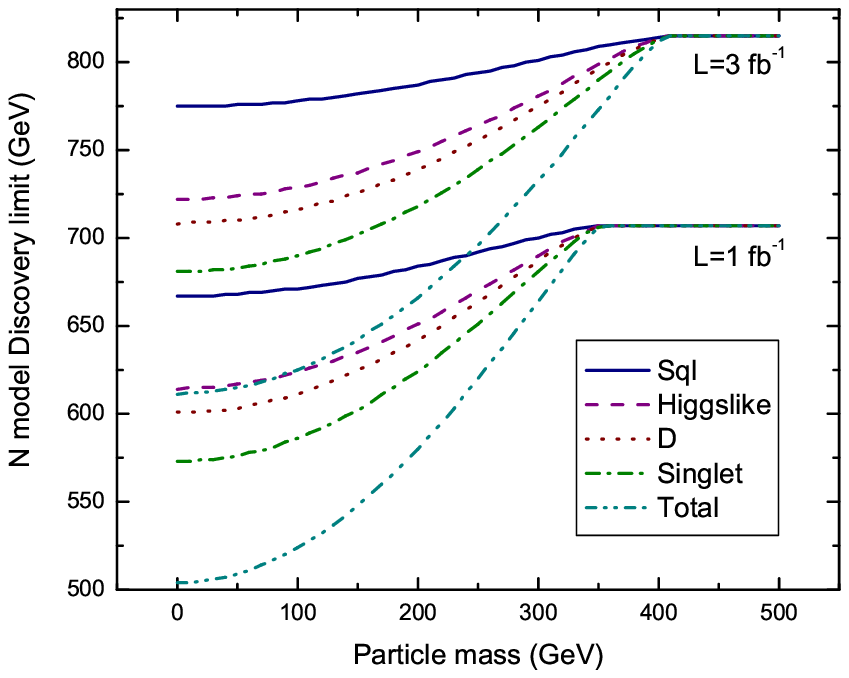}
\includegraphics[height=2.5 in]{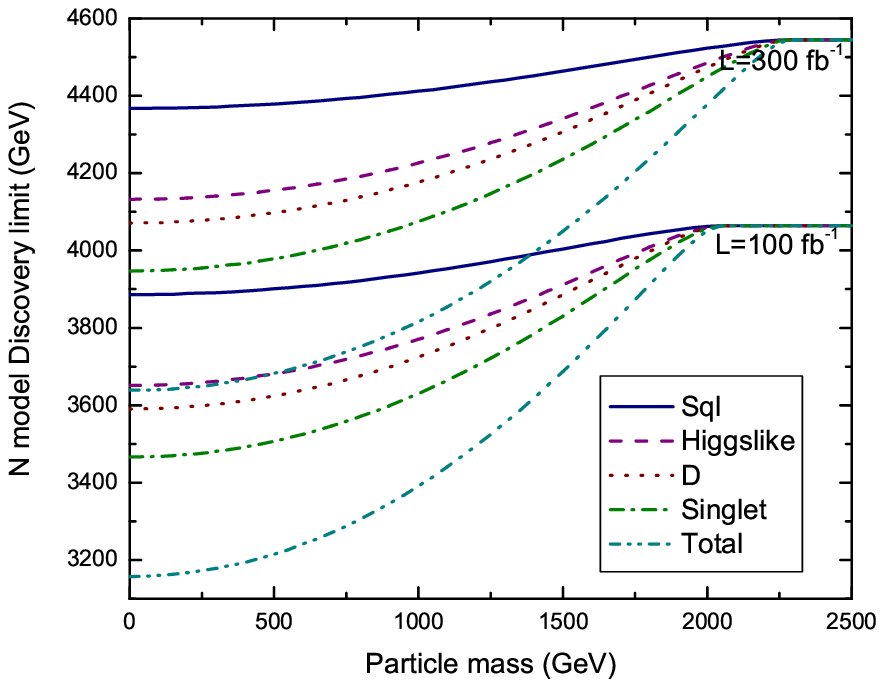}
\label{E6N}
\end{figure}

In Figs. \ref{Echi}-\ref{E6N}, we show the
discovery limit versus the non-SM particle masses in four typical
$E_6$ models for the Tevatron. For each case
only SM particles and the non-SM particles of a given type are
included. The line labelled {\em Total} includes every kind of non-SM
particle. The minimum value in that line is the extreme case in
which all particles are massless.

\section{The $Z'$ Discovery Limit in an $E_6$ Model With a Secluded Sector}
\label{secluded}
To see the effect of a definite
mass spectrum and mass matrix, we consider a concrete model
with a definite parameter set.

A supersymmetric $E_6$ model with a weakly coupled (secluded) sector
with a greatly enhanced possibility of electroweak baryogenesis
was proposed in~\cite{EWPT}. We will calculate the \zpr\ discovery
limit for that model for the parameters in~\cite{EWPT}. We first
give a short review of that model.

\subsection{Review of the Model}

There are one pair of Higgs doublets $H_u$ and $H_d$, and four SM
singlets, $S$, $S_1$, $S_2$, and $S_3$. The $U(1)'$ charges for
the Higgs fields satisfy
\begin{eqnarray}
\label {qcharge} {Q_S=-Q_{S_1} =-Q_{S_2} ={1\over 2} Q_{S_3} ~,~
Q_{H_d}+Q_{H_u}+Q_S=0 ~.~\,}
\end{eqnarray}
The superpotential for the Higgs is
\begin{eqnarray}
W_{H} &=& h S H_d H_u + \lambda S_1 S_2 S_3 ~,~\,
\end{eqnarray}
where the Yukawa couplings $h$ and $\lambda$ are respectively
associated with the effective $\mu$ term and with the secluded
sector. The existence of a number of SM singlets and the
non-diagonal nature of the superpotential are in part motivated by
explicit superstring constructions. There is an almost $F$ and $D$
flat direction involving the $S_i$, with the flatness lifted by a
small Yukawa coupling $\lambda$. For a sufficiently small value of
$\lambda$, the \zpr \ mass can be arbitrarily large. For example,
if $ h\sim 10 \lambda$, one can generate a $Z-Z'$ mass hierarchy
in which the $Z'$ mass is of order 1 TeV.

Because the representations of $E_6$ are anomaly free, we consider
the three families of SM fermions, one pair of the Higgs doublets
($H_u$ and $H_d$) from three ${\bf 27}$s, and a number of SM
singlets, exotics, and additional Higgs-like doublets. The
embedding of  the SM fermions is obvious, and we assume that the
Higgs doublets ($H_u$ and $H_d$) are the doublets in ${\bf 10}$
(or ${\bf 5}$ and ${\bf {\bar 5}}$) in the third ${\bf 27}$. In
addition, we assume that the four SM singlets $S$, $S_1$, $S_2$,
$S_3$ are the $S_L$, $S_L^*$, $S_L^*$ and ${\overline N}^*$
respectively in two pairs of ${\bf 27}$ and $\overline{{\bf 27}}$.
We include the extra $S_L$ and $\overline{N}$ so that there are three
complete pairs from ${\bf 27}+\overline{{\bf 27}}$ to avoid anomalies. An $H_u$ and
$\overline {H}_u$ pair from ${\bf 27}+\overline{{\bf 27}}$ is also
introduced for gauge unification. For simplicity, we assume that
the other particles in the two pairs of ${\bf 27}$ and
$\overline{{\bf 27}}$ are absent or very heavy.

From $Q_S = {1 \over 2} {Q_S}_3$, we obtain
\begin{eqnarray}
 \tan \theta &=& {\sqrt{15} \over 9} ~.~\,
\end{eqnarray}
The $U(1)'$ charges for the Standard Model fermions and exotic
particles are given in Table \ref{E6charge}. The general
superpotential and soft terms are given in \cite{ZZ,EWPT}.

\subsection{The $Z'$ Discovery Limit With A Definite Parameter Set}

To discuss the effect of non-SM particles on the \zpr\ discovery
limit, we need to specify their mass matrices. We use a specific
set of typical Yukawa couplings and soft terms. These were an
example of a set which leads to a strong enough first order phase
transition for electroweak baryogenesis\footnote{The parameters
we use are slightly different from those used in
\cite{EWPT}, since here we only use the tree level mass matrix, while in
\cite{EWPT} the spectrum is obtained through the full one-loop effective
potential. However, the basic features of the strong first order phase
transition will not be affected.}. The mass spectrum for the
various particles are listed in Tables
\ref{Higgsmass},~\ref{neumass} and \ref{Exomass}.

\begin{table}[t]
\caption{The CP even and CP odd Higgs boson masses in GeV at tree
level. The light masses are mainly $SU(2)$ singlets and are
consistent with experimental limits~\cite{HLM}. }\vspace{0.4cm}
\begin{center}
\begin{tabular}{|c|c|c|c|c|c|c|c|c|c|}
\hline $H_1^0$ & $H_2^0$ & $H_3^0$ & $H_4^0$ & $H_5^0$ & $H_6^0$ &
$A_1^0$ & $A_2^0$ & $A_3^0$ & $A_4^0$  \\
\hline  101 & 150 & 151 & 169 & 229 & 931 & 3
& 62 & 261 & 282\\
\hline
\end{tabular}
\end{center}
\label{Higgsmass}
\end{table}

\begin{table}[t]
\caption{The chargino and neutralino masses in GeV at tree level.}
\vspace{0.4cm}
\begin{center}
\begin{tabular}{|c|c|c|c|c|c|c|c|c|c|c|}
\hline  $\tilde{\chi}_1^{\pm}$ & $\tilde{\chi}_2^{\pm}$ &
$\tilde{\chi}_1^0$ & $\tilde{\chi}_2^0$ & $\tilde{\chi}_3^0$ &
$\tilde{\chi}_4^0$ & $\tilde{\chi}_5^0$ &
$\tilde{\chi}_6^0$ & $\tilde{\chi}_7^0$ & $\tilde{\chi}_8^0$ & $\tilde{\chi}_9^0$  \\
\hline  480 &105 & 84 & 106 & 159 & 213 & 225 &
228 & 452& 876 & 990 \\
\hline
\end{tabular}
\end{center}
\label{neumass}
\end{table}

\begin{table}[t]
\caption{Typical squark, slepton and exotic particle
(fermion/boson) masses in GeV at tree level.}\vspace{0.4cm}
\begin{center}
\begin{tabular}{|c|c|c|c|c|}
\hline squark & slepton & HiggsExo(f/b) & QuarkExo(f/b) &
SingletExo(f/b) \\
\hline  336 & 336 & 180/358 & 180/358 & 180/358 \\
\hline
\end{tabular}
\end{center}
\label{Exomass}
\end{table}

\begin{figure}
\begin{center}
\caption{The $Z'$ Discovery limit for the secluded sector model at
the Tevatron. Different classes of particles are added one after
the other. The final theoretical line includes all particles.
Lines $Neutralino$ and $Chargino$ are almost on top of each
other.} \vspace{0.4cm}
\includegraphics[height=3.6in]{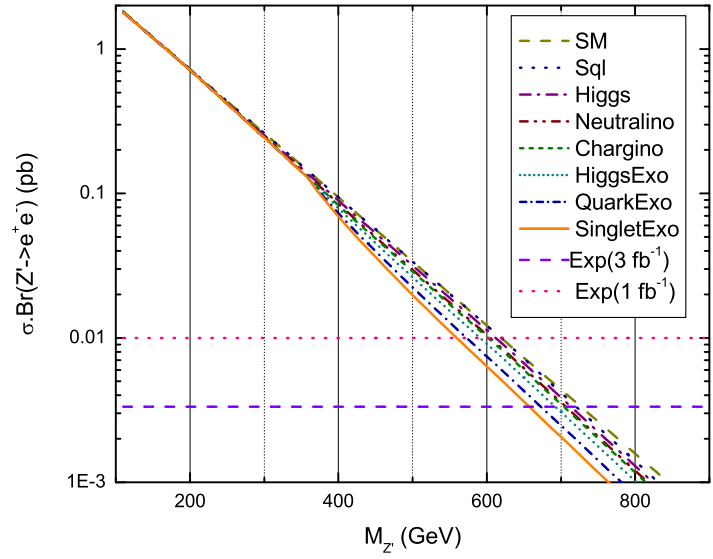}
\label{ourm}
\end{center}
\end{figure}

\begin{figure}
\begin{center}
\caption{The $Z'$ Discovery limit for the secluded sector model at
the LHC. Lines $Neutralino$ and $Chargino$ are almost on top of
each other.} \vspace{0.4cm}
\includegraphics[height=3.6in]{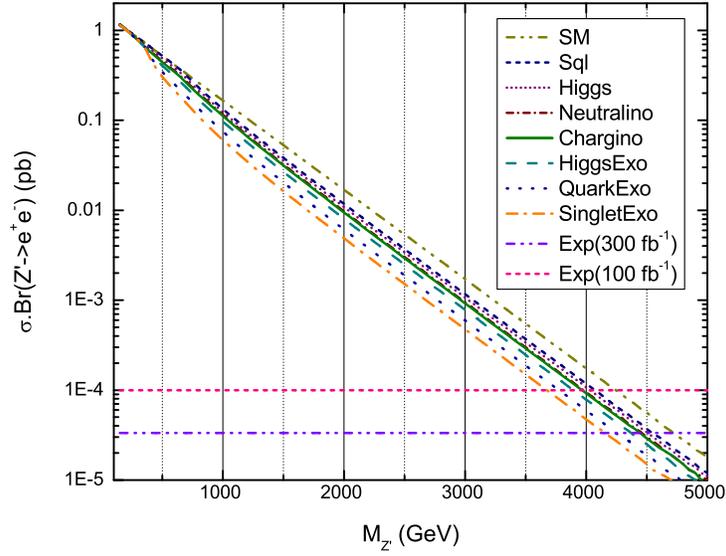}
\label{ourmlhc}
\end{center}
\end{figure}

We classify the non-standard model particles into 7 classes. The
superpartners of standard model quarks and leptons, the Higgs
fields, the neutralinos, the charginos, the Higgs-like exotics,
the quark-like exotics, and the singlet exotics.  To show their
contribution to the discovery limit, we include them one by one
into the $Z'$ decay width, in Fig. \ref{ourm} for the Tevatron and
Fig. \ref{ourmlhc} for the LHC. The line labelled $Sql$
corresponds to including the $Z'$ decay into SM fermions and their
supersymmetric partners, while other non-SM particles remain too
massive to be kinematically allowed. The other lines correspond to
adding other non-SM particles one after another. The more
particles included, the lower the discovery limit, which can
become as small as 560(658) GeV at the Tevatron for $L=1~(3)
fb^{-1}$ and 3676(4154) GeV at the LHC for $L=100~(300) fb^{-1}$.
The intersection points are given in Table \ref{ourml} and Table
\ref{ourm2}.

\begin{figure}
\caption{The $Z'$ discover limit for the secluded sector $E_6$
model at the Tevatron and LHC.} \vspace{0.4cm}
\includegraphics[height=2.5 in]{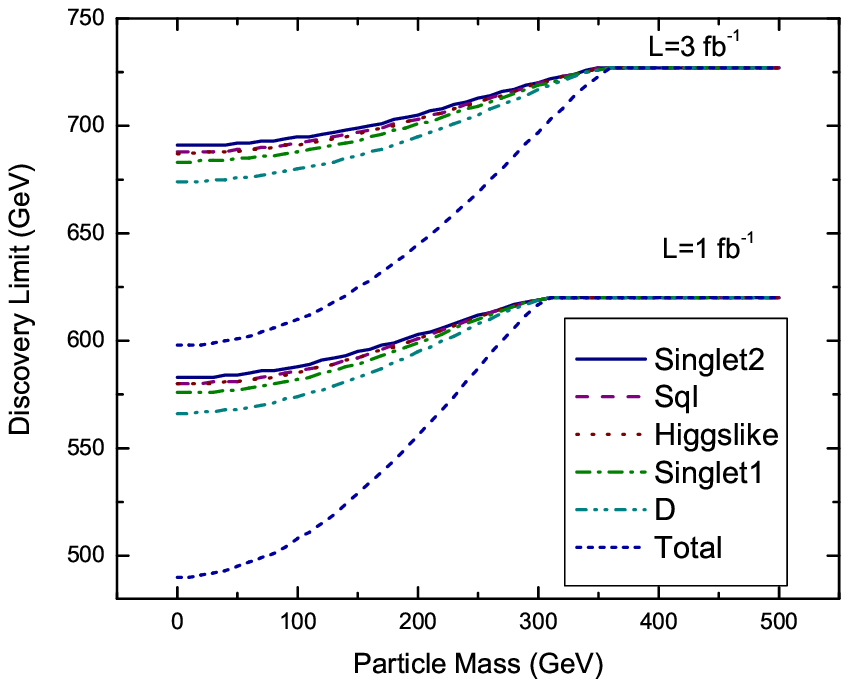}
\includegraphics[height=2.5 in]{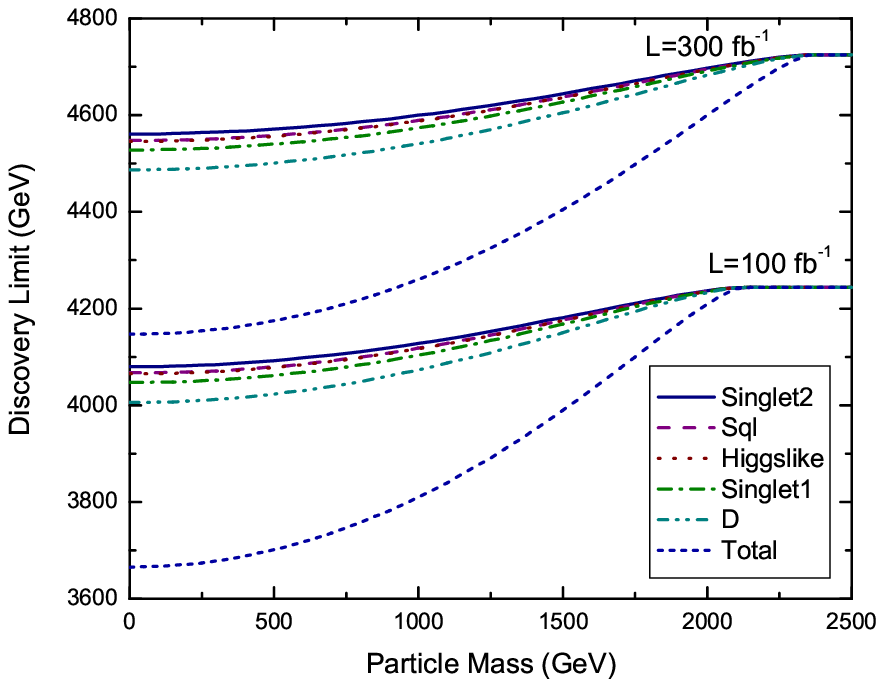}
\label{MLV}
\end{figure}

As in the generic discussion of $E_6$ models, we can vary the
non-SM particle masses to see how they affect the discovery limit,
as shown in Fig. \ref{MLV}. The only difference is that in this
model, we separate the singlets into two groups, $Singlet1$
represents the singlets from $3\times {\bf 27}$, and $Singlet2$ is for
the singlets belonging to ${\bf 27}+\overline{{\bf 27}}$.
\begin{table}[t]
\caption{ The $Z'$ Discovery limit (GeV) for the secluded sector
model at the Tevatron. In each column an additional type of decay
channel in included. \label{ourml}}
\begin{center}
\begin{tabular}{|c|c| c| c| c| c| c| c| c|}
\hline fields & SM & Sql & Higgs
& Neutralino & Chargino & HiggsExo & QuarkExo& SingletExo \\
\hline $L=1 fb^{-1}$ & 619& 620& 612& 605 & 602& 591& 573& 560\\
\hline $L=3 fb^{-1}$ & 727& 721& 714& 707 & 704& 693& 674& 658\\
\hline
\end{tabular}
\end{center}
\end{table}
\begin{table}[t]
\caption{Same as Table \ref{ourml}, for the LHC.\label{ourm2}}
\begin{center}
\begin{tabular}{|c| c| c| c| c| c| c| c|c|}
\hline fields & SM & Sql & Higgs
& Neutralino & Chargino & HiggsExo & QuarkExo& SingletExo \\
\hline $L=100 fb^{-1}$ & 4244& 4072& 4032& 3977 & 3965& 3894& 3774& 3676\\
\hline $L=300 fb^{-1}$ & 4725& 4551& 4511& 4456 & 4444& 4372& 4252& 4154\\
\hline
\end{tabular}
\end{center}
\end{table}

\section{Conclusions and Discussions}
In this paper, we discussed how the Higgs, exotic particles, and
supersymmetric partners would affect the $Z'$ discovery limit in
$E_6$ models. We first considered two limiting cases, $Z'$ decays
only into standard model particles and $Z'$ decays into all
particles with all particles massless. We showed how the total
decay width and branching ratios into fermions would differ in
these two limiting case, typically by a factor of 5 to 10.
We then studied the discovery limits of
different $E_6$ models at the Tevatron and LHC. We showed how the
discovery limits would be affected in the two limiting cases and as we
varied the masses of the exotic particles and sparticles.
The discovery limits could be lowered by up to $\sim$ 200 GeV at the Tevatron and
up to $\sim$ 1 TeV at the LHC.
As a concrete example, we considered the mass spectrum for  a typical
set of parameters
for the exotics and sparticles
for an $E_6$ model with a secluded sector.

\section*{Acknowledgments}
This research was supported in
part by the U.S.~Department of Energy under
 Grant No.~DOE-EY-76-02-3071.


\end{document}